\documentclass[aps,amsmath,amsfonts,twocolumn]{revtex4-1}%
\usepackage{mathtools}
\usepackage{verbatim}
\usepackage{hyperref}
\usepackage{epsfig,color,amsfonts,amsmath,amsthm,comment,natbib}
\usepackage{color}
\usepackage{graphicx}
\usepackage{braket}
\usepackage{amsthm}
\usepackage{amssymb}
\usepackage{amsmath}
\usepackage{ulem}
\usepackage{tikz}
\usepackage{subcaption}

\DeclareMathOperator{\Tr}{Tr}

\newcommand{\integers}{\mathbb{Z}}

\newcommand{\IK}{\color{red}}

\newcommand{\MW}{\color{blue}}

\newcommand{\old}{\color{black}}

\begin{document}

\title{ Fragmentation and Novel Prethermal Dynamical Phases in Disordered, Strongly-Interacting Floquet Systems}

\author{Matthew Wampler$^{1}$ and Israel Klich$^{1}$}
\affiliation{$^1$Department of Physics, University of Virginia, Charlottesville, Virginia 22903, USA
}

\begin{abstract}
We explore how disorder and interactions conspire in lattice models with sequentially activated hopping to produce novel k-body (or many-body) localized phases. Specifically, we show that when disorder is added to the set of interacting floquet models considered in [Wampler and Klich arXiv:2209.09180], regions in parameter space near the special points where classical-like dynamics emerge are stabilized prethermally (or via many-body localization in some cases) producing new families of interesting phases.  We also find that this disordered system exhibits novel phases in regions of parameter space away from the special, Diophantine points.  Furthermore, the regions in parameter space where Hilbert space fragmentation occurs in the clean system (leading to Krylov subspaces exhibiting frozen dynamics, cellular automation, and subspaces exhibiting signs of ergodic behavior) may also be stabilized by the addition of disorder.  This leads to the emergence of exotic dynamics within the Krylov subspace. 
\end{abstract}


\maketitle
\section{Introduction}
Periodic driving of quantum systems has emerged as an exciting tool that may be used to engineer otherwise exotic behavior \cite{Rudner2020FTI}.  Furthermore, periodically driven (Floquet) systems may support phases that are forbidden in systems evolving under static Hamiltonians.  Two prominent examples are Discrete Time Crystals \cite{Else2020DTCRev,Sacha2020DTCBook} and anomalous Floquet topological insulators \cite{titum2016anomalous,Nathan2019AFI,Nathan2021AFI}.  Time Crystals are a proposed phase \cite{Wilczek2012TimeCrystal} in which continuous time translation symmetry of a system is spontaneously broken (in analogy with the spontaneous breaking of spatial translation symmetry in the formation of crystal lattices).  Following a No-Go theorem for time crystals in static systems \cite{watanabe2015notimecrystal}, it was discovered that it is possible for the discrete time translation symmetry in periodically driven systems to be spontaneously broken forming Discrete Time Crystals \cite{Else2016DTC}.     Anomalous Floquet topological insulators take advantage of the inherently periodic nature of the non-interacting quasi-energy spectrum of periodically driven systems to exhibit novel topological features in the band structure that are impossible for static systems.  This anomalous band structure was realized in a model by Rudner-Lindner-Berg-Levin (RLBL) \cite{rudner2013anomalous}.  By adding a disordered on-site potential to the RLBL model, it was then found that the system supports a robust, new topological phase called the anomalous Floquet-Anderson topological insulator \cite{titum2016anomalous}.  The physical manifestation of this exotic, topological band structure is the emergence of chiral edge modes existing alongside a fully localized bulk.  Both Discrete Time Crystals and the anomalous topological edge behavior of anomalous Floquet topological insulators have been realized across a variety of physical platforms \cite{Zhang2017DTCExp,Frey2022DTCexp,Peng2016AFAIExp,Maczewsky2017AFAIExp,Mukherjee2017AFAIExp,Wintersperger2020AFAIExp}.            

A priori, it may be surprising that Floquet systems may exhibit robust phases since energy may be indefinitely absorbed from the drive, eventually leading to a featureless, infinite temperature state \cite{Lazarides2014Therm,Dalessio2014Therm,Ponte2015Therm}.  However, this thermalization may be combated using three main mechanisms:  1) The driven system is connected to a reservoir, which acts as a heat sink, leading to non-trivial non-equilibrium steady states \cite{Dehghani2014Dissipative,Iadecola2015Bath,Iadecola2015Bath2,Seetharam2015Bath}.  2) Only systems where energy is absorbed exponentially slowly from the drive are considered, leading to a pseudo-stable ``prethermal'' phase \cite{Canovi2016pretherm,Mori2016pretherm,Kuwahara2016pretherm,Abanin2017pretherm1,Abanin2017pretherm2}.  3) Disorder is added to the system, resulting in a localizing effect, that prevents thermalization.  This phenomena is referred to as many-body localization (MBL) \cite{Abanin2019MBLRev,Fleishman1980MBL,DAlessio2013MBL,Ponte2015MBL,Ponte2015MBLPRL,Lazarides2015MBL,Imbrie2016MBL,Abanin2016MBLFloquet,Khemani2016MBL,Agarwala2017MBL} and is an interacting generalization of Anderson localization \cite{Anderson1958localize}.

In constrained systems, there exists yet another route towards ergodicity breaking - Hilbert Space Fragmentation \cite{Sala2020HFrag,Moudgalya2022Scars,Motrunich22}.  In this case, the full Hilbert space is broken into subspaces that evolve independently.  This leads to cases where a system may have some Krylov subspaces that thermalize while others do not.  When the size of the non-thermal Krylov subspaces scales polynomially with system size (i.e. only representing a measure zero portion of the full Hilbert space), the states in these subspaces are referred to as quantum many-body scars \cite{Turner2018Scars,Ho2019Scars,Moudgalya2022Scars}.

In this work, we consider a broad class of Floquet models where hoppings between neighboring pairs of sites are sequentially activated.  A large number of Floquet systems that have received theoretical and/or experimental attention are contained within this class of models (e.g. \cite{rudner2013anomalous,titum2016anomalous,Nathan2019AFI,Nathan2021AFI,Kumar2018evenodd,friedman2019integrable,Ljubotina2019evenodd,Piroli2020evenodd,Lu2022EvenOddPRL}).  A recent investigation \cite{wampler2022arrested} found that the dynamics of clean, interacting systems in this class of models may become exactly solvable for certain driving frequencies and interaction strengths.  Specifically, these parameter values lead to evolution of Fock states into Fock states.  The special points in parameter space where this occurs are found by solving an emergent set of Diophantine equations \cite{Cohen2007Dioph}.  At other points in parameter space (also found via a set of Diophantine equations), the Hilbert space is fragmented into subspaces supporting either Frozen dynamics, classical cellular automation \cite{Wolfram1983CA}, or ergodic behavior.  

Here, we add a disordered potential to the class of interacting, Floquet systems considered in \cite{wampler2022arrested}.  We find that the disorder stabilizes, via K-body localization (described below) \cite{Aizenman2009kbl}, the dynamics of systems perturbed away from the special, Diophantine points in parameter space, leading to novel, robust phases.  The exotic dynamics of these phases may include, for example, the spontaneous breaking of time translation symmetry to form Discrete Time Crystals.  

Furthermore, we find that there are other regions in parameter space, away from any special, Diophantine points, that also represent K-body localized phases.  These regions are given by values of interaction strength and driving frequency that `almost' (see Sec. \ref{Sec: slowed dynamics}) satisfy Diophantine conditions.  In addition, at the points in the clean model where Hilbert space fragmentation occurs, the added disorder ensures that the frozen and cellular automation Krylov subspaces are stable to perturbations (in driving frequency and interaction strength) away from the special points in parameter space.  In some cases the subspace is localized by the disorder.  In other cases, the cellular automation dynamics of the subspace is stabilized over long time scales but is eventually expected to thermalize.        

Note, the stability of our results hinge on K-body localization instead of the full many-body localization.  K-body localization is a generalization of MBL where a system containing up to a maximum number of particles, K, is localized by disorder (thus MBL is given in the limit of $K\rightarrow\infty$).  Unlike MBL, which has only been rigorously established in one dimension, K-body localization is established in generic dimensional systems \cite{Aizenman2009kbl}.  However, K-body localized systems containing more then K particles will eventually be thermalized via $K+1$ particle correlations.  Thus, in the thermodynamic limit, we expect our results describe the system prethermally (except for cases, especially in 1D, where full MBL may occur). 

To help illustrate our results throughout this work, we will use a Hubbard interacting RLBL-like model on a square lattice.  In addition to the model being particularly clear for expository purposes, it has also been the center of recent interest in \cite{Nathan2021AFI} where it was found that the model supports a novel topological phase called a correlation-induced anomalous Floquet insulator (CIAFI).  The phase is characterized by a Hierarchy of topological invariants and supports quantized magnetization density.  We describe how these results may be viewed from the perspective of this work and describe new insights into the system that the Diophantine framework provides.  

The rest of this paper is structured as follows.  In Section \ref{Section:Scipost review}, we briefly review how Diophantine equations emerge in clean, periodically driven systems and their implications for the dynamics at special driving frequencies and interaction strengths (as described in \cite{wampler2022arrested}).  In Section \ref{Sec: slowed dynamics}, we perturbatively describe the evolution of these (so far clean) systems with parameter values close to the special, Diophantine points.  Section \ref{sec:classical ev disorder} describes how, once disorder is added, the evolution in this perturbative regime becomes K-body localized.  For the example case of the Hubbard-RLBL model, we provide a phase diagram for where this localization occurs.  Section \ref{sec: subspaces} describes the stability of subspaces when Hilbert space fragmentation is weakly broken by perturbing away from points where a few (but not all) of the conditions for Fock state to Fock state evolution are satisfied.  In Section \ref{section: numerics}, we corroborate the above results with numerical evidence.  Finally, in Section \ref{sec: conclusion} we provide concluding remarks.

\section{Review: Interacting Floquet models, Diophantine Equations, and Hilbert Space Fragmentation}
\label{Section:Scipost review}

Following \cite{wampler2022arrested}, we look for conditions on fermion models to evolve Fock states into Fock states deterministically.  We consider periodically-driven models where hopping between neighboring sites are sequentially activated.  Namely, we divide the period, $T$, of the Floquet drive into $M$ steps where, during step $m$, particles are only allowed to hop between pairs of sites given by a set $A_m$.  Interactions are then added to this free-hopping evolution, but we restrict ourselves to interactions that do not contain terms connecting two (or more) of the otherwise disjoint pairs with activated hopping.  Specifically, evolution during the Floquet period, $U(T)$, is given by

\begin{eqnarray} \label{eq: floquet steps}
U(T)= U_M ...U_2 U_1 
\end{eqnarray}
where $U_m=e^{-i \tau {\cal H}_m }$, $\tau = \frac{T}{M}$, and
\begin{gather}
    {\cal H}_m = -t_{hop}\sum_{(i,j)\in A_m;\sigma}  (a_{i,\sigma}^\dagger  a_{j,\sigma} + h.c.) + {\cal H}_{\text{int}} (\mathbf{V}_m).
    \label{eq: RLBL Hamiltonian}
\end{gather}
where $\mathbf{V}_m$ is a set of interaction parameters.  For the rest of this work, we will set $t_{\text{hop}}=1$ unless otherwise noted.

\begin{figure}[h]
    \centering
    \includegraphics[width=0.4\textwidth]{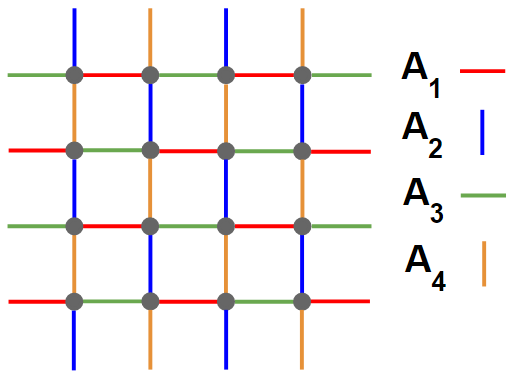}
    \caption{The RLBL model. Hopping is sequentially activated among neighbouring sites connected by the set $A_m$, $m=1,...,4$.}
    \label{fig:RLBL}
\end{figure}

As an example, consider the case of the RLBL model with Hubbard interactions.  In this case, we set $M=4$, choose $A_m$ as given in Fig. \ref{fig:RLBL}, and set

\begin{gather}
    {\cal H}^{\text{Hubbard}}_{\text{int}}(V) = V \sum_{i} n_{i,\uparrow} n_{i,\downarrow}
\end{gather}
with $n_{i,\sigma}=a_{i,\sigma}^\dagger  a_{i,\sigma}$.  Note, the Hubbard interaction is on-site and thus leaves the pairs connected by $A_m$ disjoint.

The next step is to find conditions for when individual, activated hopping pairs map Fock states into Fock states.  Since the site pairs are disjoint, we may do this individually for each pair during each step $m$ of the evolution.  

In the Hubbard-RLBL model a 2-site pair has 16 possible initial Fock states.  Let us first consider the case of a single spin up at one site with an empty neighbor. We can ignore the interacting term, and compute directly the probability, $p$, for the particle to hop to site $2$. We have $$p=|\bra{vac}a_2 e^{i \tau (a_1^{\dag}a_2+a_2^{\dag}a_1)}a_1^{\dag}\ket{vac}|^2=\sin^2{\tau},$$ 
showing that when $\tau=\frac{\pi}{2} \ell$ with $\ell \in \integers$, the evolution maps this particular Fock state to a Fock state.  Namely, the particle will remain at its initial site for $\tau$ with $\ell$ even and hop to the neighboring site when $\ell$ is odd.  We repeat this procedure for the other $15$ possible initial Fock states. The lines in the following table summarize the  conditions we find on $\tau$ and $V$ and the resulting type of evolution:
\begin{gather} \label{eq: freeze condition}
 \begin{tabular}{c | c | c}
    particles & $\tau$ & $V$ \\ \hline
      1 or 3, frozen & $ \frac{\pi}{2} \ell$, $\ell$  even &  arbitrary\\     1 or 3, swap & $  \frac{\pi}{2} \ell$, $\ell$  odd &  arbitrary\\
      2, opposite spins, frozen & $ \frac{\pi}{2} \sqrt{2 m n - n^2}$, n even & $ \frac{4 (n-m)}{\sqrt{2 m n - n^2}}$ \\ 
       2, opposite spins, swap & $ \frac{\pi}{2} \sqrt{2 m n - n^2}$, n odd & $ \frac{4 (n-m)}{\sqrt{2 m n - n^2}}$ \\
      otherwise & any & any
 \end{tabular}
\end{gather}
where $\ell,m,n \in \integers$, and $2mn-n^2 > 0$.  
The left column refers to the number of particles in the initial state of the 2-site pair (spin-up + spin-down) and the type of evolution we get.
Thus, for example, if $n$ is odd, and we start with a an up/down pair (a doublon) sitting at site $1$, the doublon will hop to site $2$. On the other hand, if $n$ is even, the doublon will stay at site $1$. 

For a generic Fock state in the full system to evolve to another Fock state, we need every 2-site activated pair to evolve deterministically.  Hence, we require all the conditions in Eq. \eqref{eq: freeze condition} to be satisfied simultaneously.  This leads to the following restriction on $\ell,m,n$

\begin{gather} \label{eq: diophantine Floquet Hubbard}
    \ell^2 + n^2 = 2 m n.
\end{gather}

The equation \eqref{eq: diophantine Floquet Hubbard}, where we are only interested in integer solutions for $\ell,m,n$, is a  Diophantine equation \cite{Cohen2007Dioph}.
Diophantine equations are an active area of mathematical research and, contrary to their often innocuous appeareance, there only exists general solution methods for a few special classes of equations.  

For the class of interacting Floquet models considered in this work, the Diophantine equations that emerge from the conditions of Fock state to Fock state evolution may not, in general, be solvable.  Fortunately, in the Hubbard-RLBL case, Eq. \eqref{eq: diophantine Floquet Hubbard} is a polynomial Diophantine equation of degree 2 for which general solution methods exist.  The solutions can be parameterized as 

\begin{subequations} \label{eq: diophantine solution}
\begin{gather}
    \ell = d(w_1^2 - w_2^2) \\
    m = d(w_1^2 + w_2^2) \\
    n = d(w_1 + w_2)^2
\end{gather}
\end{subequations}
where $w_1, w_2 \in \integers$, $w_1,w_2$ are coprime, and $d\in \frac{1}{\xi}\integers$ with $\xi=gcd((w_1^2 - w_2^2),(w_1^2 + w_2^2),(w_1 + w_2)^2)$.  

Thus, the following values of $\tau,V$ yield deterministic Fock state to Fock state evolution in the Hubbard-RLBL model

\begin{eqnarray}
\tau=\frac{\pi}{2} d(w_1^2 - w_2^2)\,\,;\,\, V=\frac{8 w_1 w_2}{|w_1^2-w_2^2|}.
\label{eq: tau and V hubbard}
\end{eqnarray}

Importantly, note that the analysis leading to Eq. \eqref{eq: tau and V hubbard} was independent of the fact that the driving procedure was RLBL.  Thus, any periodic drive with sequentially activated hopping pairs, in any dimension, with Hubbard interactions will also exhibit deterministic Fock state to Fock state evolution at the special points in parameter space given by Eq. \eqref{eq: tau and V hubbard}.

In summary, for each step $m$ in the Floquet evolution, the evolution of a special, Diophantine point given by an evolution time $\tau_{0}$ and a set of interaction parameters $\mathbf{V}_{0,m}$ is given by

\begin{gather}
    U_m(\mathbf{V}_{0,m},\tau_{0}) = {\cal P}_m
\end{gather}
where ${\cal P}_m$ is a (complex) permutation matrix on Fock states \footnote{A complex permutation matrix is a matrix where every row and column has a single non-zero element whose modulus is $1$.}.  Furthermore, for local interactions, ${\cal P}_m$ deterministically updates the occupation of individual sites based on the occupation of neighboring sites.  Starting from a product state, the evolution under an operator such as this, up to a phase, can be thought of as a classical cellular automation. 

What happens if only some of the conditions for Fock state to Fock state evolution are satisfied?  In this case, the Hilbert space will fragment into subspaces.  States in some subspaces will still evolve under cellular automation, while states in other subspaces are, in general, expected to ergodically explore their subspace.

For example, consider a Hubbard-Floquet model with a generic sequentially activated hopping.  Like the Hubbard-RLBL model, the conditions for Fock state to Fock state evolution in this model are given by \eqref{eq: freeze condition} (see discussion after Eq. \eqref{eq: tau and V hubbard}).  Now suppose only the fourth condition in \eqref{eq: freeze condition} is satisfied (for example 
 $n=3,m=2$ with $\tau=\frac{\sqrt{3}}{2}\pi$ and $V=\frac{4 \sqrt{3}}{3}$).  This condition will state that an up down pair at neighbouring sites will swap spins. Therefore, the subspace of states with exactly one particle on each site (though the spin of each particle is left generic) is invariant under the evolution.  Evolving any state in this subspace by $U_m$ will still be equivalent to evolving it by ${\cal P}_m$ since there are no 2-site activated pairs with 1 or 3 particles in the system.  This implies that this exponentially large subspace will evolve as a classical  process of spin swaps.  On the other hand, Fock states that do have 2-site pairs with 1 or 3 particles will evolve into superpositions of Fock states under $U$.  For general hopping activation sequences, this leads to an ergodic exploration of the complimentary subspace.  The full Hilbert space is thus fragmented into independent subspaces exhibiting either cellular automation or ergodic evolution.    

\section{Quantum dynamics in slow motion}
\label{Sec: slowed dynamics}

We now investigate how the systems considered in the previous section evolve when parameter values are perturbed away from the special, Diophantine points.  Now the evolution generates super-positions of Fock states and therefore entanglement. In this case, the evolution during each Floquet period is given by the cellular automation of the special point times an evolution with an effective local Hamiltonian during a reduced time compared with the original evolution period.  In other words, the correction to the classical cellular automation is a `slow motion' quantum dynamics.     

\subsection{Perturbation in Time}

We first consider a perturbation in the evolution time $\tau = \tau_{0} + \delta \tau$.  We therefore have that evolution of a step $m$ is given by 

\begin{gather}
    U_m(\mathbf{V}_{0,m},\tau_{0} + \delta \tau) =U_m(\mathbf{V}_{0,m},\tau_{0})U_m(\mathbf{V}_{0,m},\delta \tau) \\
    ={\cal P}_m e^{-i \delta \tau {\cal H}_m } \label{eq: vary tau pert}
\end{gather}
Combining this with Eq. \eqref{eq: floquet steps} we have that the evolution of the full Floquet period is given by

\begin{gather}
    U = {\cal P}_M e^{-i \delta \tau {\cal H}_M } ... {\cal P}_2 e^{-i \delta \tau {\cal H}_2 } {\cal P}_1 e^{-i \delta \tau {\cal H}_1 } \\
    \equiv  {\cal P}  U_{quantum}
    \end{gather}
 where ${\cal P}={\cal P}_M ... {\cal P}_2 {\cal P}_1$ is the unperturbed permutation, and the $U_{quantum}$ is the quantum correction to the dynamics:
   \begin{gather}
   U_{quantum}= \\
 = e^{-i \delta \tau {\cal P}^{\dag}_1{\cal P}^{\dag}_2{\cal P}_{M-1}^{\dag}{\cal H}_M {\cal P}_{M-1}..{\cal P}_2 {\cal P}_1} ...  e^{-i \delta \tau {\cal P}^{\dag}_1{\cal H}_2 {\cal P}_1}  e^{-i \delta \tau {\cal H}_1 }.\nonumber
    \end{gather}
We note that the as long as the range of the permutations $\cal P$ is finite, the dynamics $U_{quantum}$ cannot build correlations further away from the range of allowed classical dynamics. Moreover, the generation of super-positions of states is now governed by the slow  time scale $\delta \tau$. 

When tracking the evolution of a cluster of initial particles when the perturbation is small ($\delta \tau t_{\text{hop}}, \delta \tau V_{0,m} \ll 1$ ),  it is convenient to think of the quantum correction as $  U_{quantum}= e^{-i \delta \tau {\cal H}_{\text{eff}} }$
where ${\cal H}_{\text{eff}}$ is given to lowest order in $\delta \tau$ as  
\begin{gather}
    {\cal H}_{\text{eff}}\simeq \sum_{m=1}^M {\cal P}_1^\dagger  {\cal P}_2^\dagger ... {\cal P}_{m-1}^\dagger {\cal H}_m {\cal P}_{m-1} ... {\cal P}_2  {\cal P}_1 .
    \label{eq: Heff}
\end{gather}
Let us again take as an example the Hubbard-RLBL model.  In this case, the vicinity of frozen dynamics is particularly appealing. For frozen dynamics, ${\cal P}_m=I$ and thus $H_{\text{eff}} \simeq H_4 + H_3 + H_2 + H_1$ simply becomes  the static Hubbard Hamiltonian on the square lattice with $V \rightarrow 4 V_0$.  This means that, to a good approximation, ${\cal H}_{Floquet}\simeq \frac{\delta \tau}{T} {\cal H}_{Hubbard}(4V_0)$ with ${\cal H}_{Floquet}$ the usual Floquet Hamiltonian defined by $U(T) = e^{-i T {\cal H}_{Floquet}}$. In other words, the stroboscopic evolution in the system is that of a slow-motion static Hubbard evolution, i.e. after $N$ evolution cycles, at time $T N$, the system will have evolved under a static Hubbard Hamiltonian (with interaction $4V_0$) for a reduced time $\delta \tau N$.  When ${\cal P}_m \neq \mathbb{I}$, the evolution is a sequence of permutations followed by slowed, modified Hubbard evolution.  Note, the modified Hubbard evolution (given by specializing \eqref{eq: Heff} to the Hubbard-RLBL case) includes hopping terms that are modified by the permutations of lattice sites while the interaction terms are unaffected (since ${\cal H}^{\text{Hubbard}}_{\text{int}}$ counts the number of doublons, which is preserved under the freezing and swapping operations generating the dynamics at the special Diophantine points). 


\subsection{Perturbation in Interaction Strength}
Now, suppose instead that we consider a perturbation in interaction parameters $\mathbf{V}_m = \mathbf{V}_{0,m} + \delta \mathbf{V}_m$.  In the case that $\tau_0 \delta V_m \ll 1$ for all $\delta V_m \in \delta \mathbf{V}_m$, we expand
\begin{gather}
    U_m(\mathbf{V}_{0,m} + \delta \mathbf{V}_m,\tau_0) \nonumber \\ \approx{\cal P}_m \left(1 - i \int_0^{\tau_0} ds e^{i s {\cal H}_m(\mathbf{V}_{0,m})} \delta{\cal H}_{\text{int}} e^{-i s {\cal H}_m(\mathbf{V}_{0,m})}\right) \nonumber \\
    \equiv {\cal P}_m e^{-i \tau_0 {\cal H}_{\text{eff},m}}
    \label{eq: vary V pert}
\end{gather}
with ${\cal H}_{\text{eff},m}$ defined in the last line.  Thus, in a similar fashion to the perturbation in $\tau$ case, we find that   

\begin{gather}
    U = {\cal P}_M ... {\cal P}_2 {\cal P}_1 e^{-i \tau_0 {\cal H}_{\text{eff}} }
    \label{eq: full U vary V}
\end{gather}
with
\begin{gather}
    {\cal H}_{\text{eff}}\simeq \sum_{m=1}^M {\cal P}_1^\dagger  {\cal P}_2^\dagger ... {\cal P}_{m-1}^\dagger {\cal H}_{\text{eff},m} {\cal P}_{m-1} ... {\cal P}_2  {\cal P}_1 .
    \label{eq: Heff V}
\end{gather}

Illustrating again with the Hubbard-RLBL model, $H_{\text{eff},m}$ may be written explicitly in terms of creation and annihilation operators by solving each disjoint 2-site pair in $\int_0^{\tau_0} ds e^{i s {\cal H}_m(\mathbf{V}_{0,m})} \delta {\cal H}_{\text{int}}  e^{-i s {\cal H}_m(\mathbf{V}_{0,m})}$ separately and then summing.  We find, to lowest order in $\tau_0 \delta V$, that
\begin{gather}
    H_{\text{eff},m} = \delta V \sum_{i\in \text{2-site pairs}} n_{\leq \text{2-part},i} \mathbf{a}_i^\dagger {\cal T}  \mathbf{a}_i n_{\leq \text{2-part},i} + n_{> \text{2-part},i}
    \label{eq: slow near V}
\end{gather}
where we have defined 
\begin{gather}
    {\cal T} = \frac{-1}{16+V_0^2} \left(\begin{tabular}{c c c c}
         $12-V_0^2$ & $-4$ & $V_0$ & $V_0$ \\
         $-4$ & $12 - V_0^2$ & $V_0$ & $V_0$ \\
         $V_0$ & $V_0$ & $4$ & $4$ \\
         $V_0$ & $V_0$ & $4$ & $4$ 
    \end{tabular} \right) \\
    \mathbf{a}_i = \left(\begin{tabular}{c c c c}
         $a_{2 \uparrow} a_{2 \downarrow}$ &  $a_{1 \uparrow} a_{1 \downarrow}$ & $a_{1 \uparrow} a_{2 \downarrow}$ & $a_{2 \uparrow} a_{1 \downarrow}$
    \end{tabular} \right)^T
\end{gather}
and $n_{> \text{2-part},i}$ projects onto the subspace with the i$^{th}$ 2-site pair having more than 2 particles, i.e.
\begin{gather}
    n_{> \text{2-part}} = 1 - n_{\leq \text{2-part}} \nonumber \\
    = \sum_{\begin{tabular}{c}
          $a, b, c \in \{1 \uparrow, 1 \downarrow, 2 \uparrow, 2 \downarrow  \}$\\
          $a \neq b \neq c$
    \end{tabular}} n_a n_b n_c .
\end{gather}
In other words, evolution under $H_{\text{eff},m}$ corresponds to correlated hopping for any 2-site pairs containing 2 particles and a $\delta V$ energy cost of having a two-site pair with 3 or more particles. 


\subsection{Away From Special Points}
Slowed effective quantum dynamics corrections to classical Fock state permutations may occur away from the vicinity of the special Diophantine points.  Here we explore other regions in parameter space, far from special points, where the conditions for Fock state to Fock state evolution are approximately satisfied.  Specifically, 
 consider the evolution at step $m$ consisting of the set $A_m$ of the activated two site pairs,  and let ${\cal H}_{(i,j)}$ be the Hamiltonian for $(i,j) \in A_m$.

Let us define:
\begin{gather}
    D_{(i,j)}=\inf_{{\cal P}_{(i,j)}} ||  e^{-i \tau {\cal H}_{(i,j)}} - {\cal P}_{(i,j)}||_{HS}\,\,;\,\,D=\max_{(i,j)}D_{(i,j)} \label{eq: D}
\end{gather}
where $||\cdot||_{HS}$ is the Hilbert-Schmidt norm of the $(i,j)$ subspace. When $D_{(i,j)}\ll 1$ we can say that the conditions for Fock state to Fock state evolution are approximately satisfied for ${\cal H}_m$ and $\tau$. 


We may now choose a ${\cal P}_{(i,j)}$ that minimizes $D_{(i,j)}$ and write
\begin{gather}
e^{-i \tau {\cal H}_{(i,j)}}   \equiv   {\cal P}_{(i,j)}  e^{-i \tau {\cal H}_{eff,(i,j)}}.
    \label{eq: slowed in pair}
\end{gather}
As before,  $e^{-i \tau {\cal H}_{eff,(i,j)}}$ is close to the identity matrix when $D$ is small, therefore the evolution corresponds to a permutation augmented with slowed quantum dynamics.  
Using Eq. \eqref{eq: slowed in pair} the evolution of the full system for step $m$ is
\begin{gather}
    U_m = \bigotimes_{(i,j) \in A_m}  e^{-i \tau {\cal H}_{(i,j)}} 
    = {\cal P}_m e^{-i \tau {\cal H}_{eff,m}} \label{eq: gen U step slow}
\end{gather}
where we have defined 
\begin{gather}
  {\cal P}_m=  \bigotimes_{(i,j) \in A_m} {\cal P}_{(i,j)}  \label{eq:P product}\\
    {\cal H}_{eff,m} = \sum_{(i,j) \in A_m} {\cal H}_{eff,(i,j)}.
\end{gather}

Equation \eqref{eq: gen U step slow} is of the general form of equations \eqref{eq: vary V pert} and \eqref{eq: vary tau pert}, and we find analogously that the evolution of the full Floquet period is given by \eqref{eq: full U vary V} and \eqref{eq: Heff V}. 

\begin{figure}[t]
    \centering
    \includegraphics[width=0.48\textwidth]{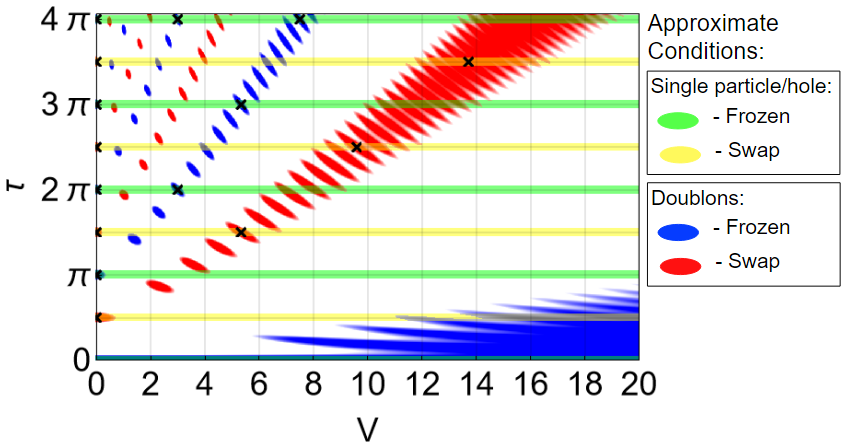}
    \caption{Parameter space regions where the Diophantine conditions \eqref{eq: freeze condition} are approximately satisfied.  Specifically, regions where the Hilbert-Schmidt norm of the difference between the evolution of a step for an activated 2-site pair and the Fock state permutation corresponding to the satisfaction of the $i^{th}$ condition in \eqref{eq: freeze condition} is less than $\epsilon=0.1$ are filled with color $i$ (as given in the legend). The special, Diophantine points where evolution is exactly a Fock state permutation are marked with a black ``$X$.''}
    \label{fig:HubbardRLBL Phase Diagram}
\end{figure}

To illustrate the appearance of slow dynamics parameter space regions away from special points, in Fig. \ref{fig:HubbardRLBL Phase Diagram} we plot regions where the Diophantine conditions are approximately satisfied in the Hubbard-RLBL model.  Namely, we plot regions where the Hilbert-Schmidt norm of the difference between the evolution of an activated pair and a SWAP or Identity permutation is less than some small cutoff.  In the figure, we consider separately when 2 site pairs with 1 or 3 particles evolve approximately as a permutation and when 2 site pairs with 2 particles of opposite spin evolve approximately as a permutation.  Parameters where pairs with 1 or 3 particles are approximately frozen (perfect swapping) are colored green (yellow) while parameters where pairs with 2 particles of opposite spin are approximately frozen (perfect swapping) are colored blue (red).  Thus, $D$ is small only in regions of overlapping colors.  Note that each special, Diophantine point is surrounded by a region of overlapping colors, but not all regions of overlapping colors contain a special point.

A couple of remarks are in order.
Note that in the Hubbard-RLBL model, special points are only found when all activated 2-site pairs are frozen or when all the pairs are perfect swapping.  This can be seen from Eq. \eqref{eq: diophantine solution} by verifying that the parity of $\ell$ and $n$ come out the same for any choice of $d,w_1,w_2$ \cite{wampler2022arrested}.  Indeed, in Fig. \ref{fig:HubbardRLBL Phase Diagram} the special points (represented by $x$) appear only when both single particle and doublon sectors are simultaneously perfectly swapping (or simultaneously perfectly frozen). However, Fig. \ref{fig:HubbardRLBL Phase Diagram} also shows that it is possible to approximately have perfect swapping in the 1 and 3 particle sector while pairs with 2 particles are approximately frozen (and vice versa).  
Another thing to note is that when only some of the Diophantine conditions are approximately satisfied, the fragmented Krylov subspaces that would emerge if the conditions were perfectly satisfied may become connected by the slowed dynamics ${\cal H}_{eff}$. 

In the next section, we add disorder to the system and find that the slowed dynamics is (in some cases) either K-body or many-body localized by the disorder.  This then stabilizes the cellular automation dynamics in regions where the conditions for Fock state to Fock state evolution are approximately satisfied leading to robust phases.

\section{Stabilizing Classical Evolution with Disorder}\label{sec:classical ev disorder}

We now add disorder to the periodically driven, interacting models considered above.  Specifically, we investigate Floquet drives of the form

\begin{eqnarray} \label{eq: floquet steps dis}
U(T)= U_{\text{dis}} U_M ...U_2 U_1 
\end{eqnarray}
where we take $U_{dis}$ to be evolution under a disordered on-site potential with no hopping, i.e. 
\begin{gather}
   U_{dis} = e^{-i \tau {\cal H}_{dis} } \label{eq: U wait}
\end{gather}
where
\begin{gather}
     {\cal H}_{dis} = \sum_{i,\sigma} v_i n_{i,\sigma}
\end{gather}
with $v_i$ uniformly distributed in $[-W,W]$.  However, the exact form of disorder doesn't play a role in the argument \footnote{Note, for example, if $H_{dis}$ is included as a constant term throughout the evolution instead of only being applied during the $U_{dis}$ portion of the drive, then all our results still hold \cite{Nathan2021AFI}.  In this case, the strength of the disorder during the first 4 steps of the floquet drive must be kept small.  The strength during the $5^{th}$, disorder only, step may be made large by lengthening the time that $U_{dis}$ is applied. }.

For Floquet systems of this type, sufficiently strong disorder will (either K-body or many-body) localize the slowed dynamics but leave the cellular automation dynamics unaffected. This happens when the cellular automation has a finite order and when the disorder is large compared to the slowed evolution.  This leads to the emergence of robust phases with stabilized cellular automation dynamics.
\subsection{Dynamics with disorder close to ${\cal P}=I$}
To illustrate how this occurs, we begin with the simpler case of when the cellular automation of the full Floquet period (though not necessarily each step $m$ of the drive) is the identity, i.e.

\begin{gather}
    {\cal P}_M ... {\cal P}_2 {\cal P}_1 \equiv {\cal P} = I 
\end{gather}
where we have defined ${\cal P}$ as the full cellular automation.  Examples of cellular automata of this type include frozen dynamics and the perfect swapping Hubbard-RLBL model with periodic boundary conditions (e.g. $\tau=\frac{\pi}{2}$, $V=0$ or $\tau=\frac{3 \pi}{2}$, $V=\frac{16}{3}$).

Using \eqref{eq: full U vary V} we find that, for parameters where the Diophantine conditions are approximately satisfied, the evolution of one Floquet period $T$ may be written

\begin{gather}
     U = U_{\text{dis}} {\cal P} e^{-i \tau {\cal H}_{\text{eff}} } \label{eq: Hdis P Heff}\\
     = U_{\text{dis}} e^{-i \tau {\cal H}_{\text{eff}} } \label{eq: Hdis then Heff}\\
    \approx  U_{\text{dis}} \left(1 - i \int_0^\tau dt e^{i t {\cal H}_{\text{dis}}} {\cal H}_{\text{eff}}(t)  e^{-i t {\cal H}_{\text{dis}}} \right) \\
     = {\cal T}e^{-i \int_0^\tau dt {\cal H}(t)}
\end{gather}
where
\begin{gather}
    {\cal H}_{\text{eff}}(t) = e^{-i t {\cal H}_{\text{dis}}} {\cal H}_{\text{eff}}  e^{i t {\cal H}_{\text{dis}}} \\
    {\cal H}(t) = {\cal H}_{\text{dis}} + {\cal H}_{\text{eff}}(t) \label{eq: H of P=I}
\end{gather}
We therefore have that the cellular automation evolution will be stable if the Hamiltonian \eqref{eq: H of P=I} is localized.

To see when this is the case, we rewrite \eqref{eq: H of P=I} as

\begin{gather}
    {\cal H}(t) = {\cal H}^{(0)} + V(t)
    \label{eq: H for MBL frozen}
\end{gather}
where  

\begin{gather}
    {\cal H}^{(0)} = {\cal H}_{\text{dis}} + \overline{{\cal H}_{\text{eff}}(t)}\\
    V(t) = {\cal H}_{\text{eff}}(t) - \overline{{\cal H}_{\text{eff}}(t)} ; \, \, \overline{V(t)} = 0.
\end{gather}
with $\overline{f(t)} = \frac{1}{\tau}\int_0^{\tau} ds f(s)$ the time average and $V(t)$ is the strictly time-dependent part of ${\cal H}(t)$.

We note that ${\cal H}_{\text{eff}}$ is a sum of local Hamiltonians sandwiched by ${\cal P}_m$ \eqref{eq: Heff V} and is thus also local.  This implies that $V(t)$ is local as well and may be written 

\begin{gather}
    V(t) = \sum_i V_i(t).
\end{gather}

Whenever ${\cal H}^{(0)}$ is MBL, we may use a theorem by Abanin, De Roeck, and Huveneers \cite{Abanin2016MBLFloquet} to show that the weak, local drive $V(t)$ will not ruin the localization of ${\cal H}^{(0)}$.  Namely, that the Hamiltonian \eqref{eq: H of P=I} will be MBL whenever

\begin{gather}
    \tau ||V_i(t)||_{HS}  \ll 1 \, \text{ and } \, \frac{\tau ||V_i(t)||_{HS}^2}{W} \ll 1 
    \label{eq: condits for Floquet local}
\end{gather}
Note that $\tau ||{\cal H}_{\text{eff}}||_{HS} \ll 1$ implies $\tau ||V_i(t)||_{HS}  \ll 1$.  Hence, for sufficiently strong disorder, the Hamiltonian \eqref{eq: H of P=I} will be MBL so long as ${\cal H}^{(0)}$ is MBL.  A corrollary of this result is that \eqref{eq: H of P=I} will be K-body localized so long as ${\cal H}^{(0)}$ is K-body localized \footnote{This is because \cite{Abanin2016MBLFloquet} uses a KAM type scheme to, order by order, find exactly the dressed $\ell$-bits of the Hamiltonian and thus prove it is MBL.  This procedure may thus be stopped at some order $K$ to show K-body localization.}.  

We have thus reduced the problem to asking whether the static Hamiltonian ${\cal H}^{(0)}$ is localized.  For 1D systems, we expect ${\cal H}^{(0)}$ to be MBL.  This is because, using a KAM type scheme, it has been shown that models of this type (i.e. Hamiltonians with a disordered term plus a weak, local interacting term) are MBL in 1D under the weak assumption of Limited Level Attraction (see \cite{Imbrie2016MBL}).  In higher dimensions, rare regions of weak disorder can cause an avalanche effect that ruins the MBL \cite{DeRoeck2017MBL,DeRoeck2017Avalanche,Potirniche2019Avalanche,Morningstar2022Avalanche}.  This delocalization, however, happens on exponentially long time scales and thus the system is prethermally localized.  Furthermore, these systems are expected to be K-body localized.  This is because the probability that the K-particle energy spectrum features the accidental resonances that ruin localization goes to zero in the thermodynamic limit \cite{Aizenman2009kbl}.

In summary, we have shown that when disorder is added to a Floquet system near a special point with ${\cal P}=I$, then the dynamics is stabilized by (many-body or K-body) localization and thus corresponds to a robust prethermal phase. We expect that as we move further away from special points, the effective evolution would resemble that of random local unitaries in which spreading has been studied in e.g.   \cite{nahum2018operator}.

\subsection{Discrete Time Crystals}
We now consider a Floquet drive that corresponds to a perfect cellular automation with some finite order $\geq1$, i.e.
\begin{gather}
    {\cal P}^{\cal O} = I; \, \, {\cal O} \in \mathbb{N} .
\end{gather}
Such dynamics, when stable to disorder and ${\cal O}>1$,  is often called a discrete time crystal. Indeed, the original time translation symmetry of the Floquet drive, $T$, has been spontaneously broken in these interacting, localized Floquet phases that now have ${\cal O} T$ time translation symmetry. 
Let us consider the evolution after ${\cal O}$ Floquet periods given by
\begin{gather}
    U^{\cal O} = \left[U_{\text{dis}} {\cal P} e^{-i \tau {\cal H}_{\text{eff}} } \right]^{\cal O} \label{eq: U^O start} \\
    = \left[U_{\text{dis}} {\cal P} \right]^{\cal O} e^{-i \tau {\cal H}_{{\cal O},\text{eff}} }
\end{gather}
where, to first order in $\tau ||{\cal H}_{eff,(i,j)} ||_{HS}$, 
\begin{gather}
    {\cal H}_{{\cal O},\text{eff}} = \sum_{a=0}^{{\cal O}-1} \left({\cal P}^\dagger U_{\text{dis}}^\dagger \right)^a {\cal H}_{\text{eff}} \left(U_{\text{dis}} {\cal P}  \right)^a.
\end{gather}

Now, note that since ${\cal P}$ is a cellular automation (and thus updates the occupancy of sites based on the occupancy of nearby sites), it transforms the number operator of a site $i$ in the following way

\begin{gather}
    {\cal P}^\dagger n_i {\cal P} = \sum_{i_1} P_{i_1} n_{i_1} + \sum_{i_1,i_2} P_{i_1 i_2} n_{i_1} n_{i_2} \nonumber \\
    + ... + \sum_{i_1,i_2,...,i_\lambda} P_{i_1 i_2 ... i_\lambda} n_{i_1} n_{i_2} ... n_{i_\lambda}\label{eq: n-transformation}
\end{gather}
where the coefficients $P_{i_1},P_{i_1 i_2}, ...,P_{i_1 i_2 ... i_\lambda}$ are only non-zero when all the sites $i_1,i_2,...,i_\lambda$ are within some finite region surrounding site $i$.  This implies that
\begin{gather}
    \left[U_{\text{dis}} {\cal P} \right]^{\cal O} e^{-i \tau {\cal H}_{{\cal O},\text{eff}} } \nonumber \\
    = {\cal P}^{\cal O} e^{-i \tau {\cal H}_{\text{loc}}} e^{-i \tau {\cal H}_{{\cal O},\text{eff}} } \nonumber \\
    = e^{-i \tau {\cal H}_{\text{loc}}} e^{-i \tau {\cal H}_{{\cal O},\text{eff}} } 
    \label{eq: Hloc then HOeff}
\end{gather}
where
\begin{gather}
    {\cal H}_{\text{loc}} = \sum_{a=0}^{{\cal O}-1} {\cal P}^a {\cal H}_{\text{dis}} {\cal P}^a .
\end{gather}
Note that $ {\cal H}_{\text{loc}} $ is a sum of local terms as in \eqref{eq: n-transformation}. 
We now repeat the steps \eqref{eq: Hdis then Heff} to \eqref{eq: H for MBL frozen} to find that

\begin{gather}
    U^{\cal O} = {\cal T}e^{-i \int_0^\tau dt {\cal H}(t)}
    \label{eq: U^O end}
\end{gather}
with 
\begin{gather}
    {\cal H}(t) = {\cal H}^{(0)} + V(t)
    \label{eq: H(t) for P finite order}
\end{gather}
and 
\begin{gather}
    {\cal H}^{(0)} = {\cal H}_{\text{loc}} + \overline{{\cal H}_{{\cal O},\text{eff}}(t)}\\
    V(t) = {\cal H}_{{\cal O},\text{eff}}(t) - \overline{{\cal H}_{{\cal O},\text{eff}}(t)} \\
     {\cal H}_{{\cal O},\text{eff}}(t) = e^{-i t {\cal H}_{\text{loc}}} {\cal H}_{{\cal O},\text{eff}}  e^{i t {\cal H}_{\text{loc}}} 
\end{gather}

Again using \cite{Abanin2016MBLFloquet}, we find that \eqref{eq: H(t) for P finite order} is k-body (many-body) localized whenever ${\cal H}^{(0)}$ is k-body (many-body) localized and \eqref{eq: condits for Floquet local} are satisfied.  Again, since ${\cal H}^{(0)}$ is a fully MBL term plus a weak, local interacting term, we expect it to be k-body (many-body) localized as discussed in the paragraph following \eqref{eq: condits for Floquet local}.  

We remark that the range of ${\cal H}_{{\cal O},\text{eff}}$, and thus $||V_i(t)||_{HS}$, increases with increasing ${\cal O}$.  By \eqref{eq: condits for Floquet local}, this implies that the region in parameter space where the system is localized will shrink rapidly for increasing ${\cal O}$, however the region will remain finite so long as ${\cal O}$ is finite.  This also suggests that general cellular automations without finite order are not likely to be stabilized by the disorder.  For example, for the perfect swapping RLBL model, the bulk cellular automation has order $1$ while the cellular automation at the edge has infinite order (since particles are transported chirally along the edge).  This is another way of viewing why the edge modes of an interacting, perfect-swapping RLBL model thermalize \cite{Nathan2019AFI,Nathan2021AFI} even while the system with periodic boundary conditions does not.      

\section{Stabilized Subspaces}
\label{sec: subspaces}
We now investigate when interacting models with sequentially activated hopping may exhibit stabilized cellular automation dynamics in Krylov subspaces even when the full Hilbert space does not support such dynamics.  Namely, we consider two main situations where this may occur. 

First, we can have all the Diophantine conditions approximately satisfied, but the corresponding cellular automation has infinite order when acting on some states (e.g edge states in the RLBL model).  However, some initial Fock states may exhibit finite orbits under the cellular automation.  These orbits may then be stabilized by disorder.  

Another possibility is when only some of the Diophantine conditions are approximately satisfied.  Here, the Hilbert space fragmentation (that occurs when a few of the Diophantine conditions are perfectly satisfied \cite{wampler2022arrested}) may be stabilized by the disorder.

In both these cases, we are thus interested in Floquet evolution that may be written
\begin{gather}
    \label{eq: U subspaces}
    U(T) = U_{\text{dis}} U_{\cal P} e^{-i \tau {\cal H}_{\text{eff}} } .
\end{gather}
Here $U_{\cal P}$ maps number states to number states only in a subspace ${\cal N}$ associated with satisfied conditions. 
Let us consider cases where 
\begin{gather}
    U_{\cal P}^{\cal O} |{\cal N}\rangle = |{\cal N}\rangle
    \label{eq: U^O in subspaces}
\end{gather}
for some finite ${\cal O} \in \mathbb{N}$.



We now ask whether the subspace ${\cal N}$ is localized under the evolution $U(T)$.

We first specialize to the situation where the subspace ${\cal N}$ is closed under the evolution of $U(T)$, i.e. 
\begin{gather}
    U(T) |{\cal N}\rangle = |{\cal N}'\rangle \,\,\, \forall \,\,\, |{\cal N}\rangle \in {\cal N}
\end{gather}
where $|{\cal N}'\rangle \in {\cal N}$.  A simple example when this (approximately) happens is when starting with a few particles far away from the edge in the RLBL model. A more elaborate example will be discussed below.

In this case, $U(T)$ is therefore block diagonal with
\begin{gather}
    U(T) = \left(
    \begin{tabular}{c c}
         $U_{{\cal N}^c}$ & $0$ \\
         $0$ & $U_{{\cal N}}$ 
    \end{tabular} \right)
\end{gather}
where $U_{{\cal N}}$ acts on the space ${\cal N}$ and $U_{{\cal N}^c}$ acts on its compliment ${\cal N}^c$.  

Now, using \eqref{eq: U subspaces} and \eqref{eq: U^O in subspaces} and  repeating the steps \eqref{eq: U^O start} through \eqref{eq: U^O end}, we have that $U_{{\cal N}}^{\cal O}$ is localized.  

We again illustrate our point using the Hubbard-RLBL model.  Consider the case where activated pairs with 1 or 3 particles are approximately perfect swapping while pairs with 2 particles of 
opposite spin are approximately frozen (i.e. regions in Fig, \ref{fig:HubbardRLBL Phase Diagram} where yellow and blue overlap).  Here, the corresponding cellular automation has infinite order.  However, the cellular automation does have finite order when acting on subspaces with a fixed, finite number of particles.  Furthermore, note that the Hubbard-RLBL evolution is $U(1)$ symmetric, thus the finite particle number subspaces are closed under $U(T)$.  Therefore, by the arguments of this section, we have that the cellular automation will be stabilized by disorder for any initial state with a fixed, finite number of particles. 

For cases where the subspace ${\cal N}$ is not closed under the evolution $U(T)$ it might be expected, in general, that the system will fail to localize.  This is because, as soon as the initial state evolves outside the subspace ${\cal N}$, the operator $U_{\cal P}$ will act to delocalize the state.  However, we investigate such cases numerically in the next section and find that initial states within ${\cal N}$ may still remain localized on prethermal time scales.      
\section{Numerical results} \label{section: numerics}

\begin{figure*}[t]
    \centering
    \includegraphics[width=\textwidth]{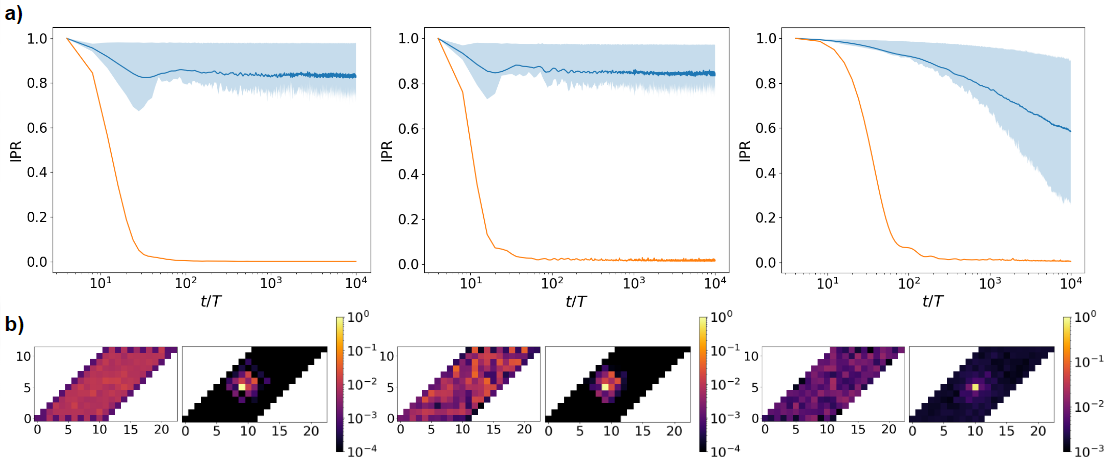}
    \caption{Localization in Hubbard-RLBL: a) Evolution of the Inverse Participation Ratio (IPR) in the Hubbard-RLBL model for a doublon initially localized in the center of the system.  From left to right, the figures correspond to the second and fourth conditions in \eqref{eq: freeze condition} being approximately satisfied (both particles and doublons are swapping, $V=\frac{16}{3}-0.05$, $\tau = \frac{3 \pi}{2} + 0.05$), the first and fourth conditions being approximately satisfied (particles are frozen and doublons are swapping, $V=7.9$, $\tau = 2 \pi + 0.1$), and solely the third condition being approximately satisfied (doublons are frozen, $V=\sqrt{2} - 0.01$, $\tau = \sqrt{2} \pi + 0.01$).  The orange curve corresponds to the evolution without disorder, while the blue curve corresponds to the evolution with disorder.  Disordered runs are averaged over $500$ realizations of $W=10$ with the range from the $25^{th}$ to the $75^{th}$ percentiles filled in with light blue.  Consistent with the theoretical arguments of Sections \ref{sec:classical ev disorder} and \ref{sec: subspaces}, the numerics suggest that disorder fully localizes the evolution in the first two cases and stabilizes the classical dynamics over exponentially long time scales in the third case.  b)  Average density of spin-up particles per site after $10,000$ driving periods from an initial doublon localized in the center of the system.  From left to right, the same three values for $V,\tau$ are used as in Fig. \ref{fig:IPR and G}a.  For each value of $V,\tau$, the densities after evolution without disorder (left) and with disorder (right) are plotted.     }
    \label{fig:IPR and G}
\end{figure*}

In this section, we numerically investigate the evolution of several example interacting Floquet systems both with and without disorder.  Namely, we investigate the stabilizing effect when disorder is added and if the evolution is consistent with localization dynamics.  

As a measure of localization, we use the Inverse Participation Ratio (IPR).  Given any state $|\Psi\rangle$, and letting $|\mathbf{n}\rangle$ be the number state basis in real space, the IPR is defined as

\begin{gather}
    \text{IPR} = \sum_{\mathbf{n}} |\langle \Psi | \mathbf{n} \rangle |^4 
\end{gather}
The IPR is $1$ for any $|\Psi \rangle$ that is a Fock state and goes as $\frac{1}{N^2}$ (where $N$ is the dimension of the Hilbert space) for an equal superposition of number states.

In Figure \ref{fig:IPR and G}a, we plot the IPR as a function of time for three example values of $V,\tau$ in the Hubbard-RLBL model starting from the initial Fock state of a doublon localized in the center of the system.  

In the first case, the second and fourth conditions in \eqref{eq: freeze condition} are approximately satisfied with $V=\frac{16}{3}-0.05$, $\tau = \frac{3 \pi}{2} + 0.05$, i.e. we have perturbed both $V,\tau$ away from the special point $V=\frac{16}{3},\tau = \frac{3 \pi}{2}$ where both single particles and doublons evolve with swapping.  Without disorder, the system will evolve with the effective slowed, interacting dynamics as discussed in Section \ref{Sec: slowed dynamics} since it is near a special point.  However, the doublon under this dynamics may still generate superpositions and spread throughout the system.  Thus, the IPR decreases.  Note, however, that the Fock state permutation at the special, Diophantine point has finite order (namely, order 1 since the perfect swapping RLBL model acts as the identity in the bulk of the system).  Therefore, by the analysis of Section \ref{sec:classical ev disorder}, it is expected that disorder will K-body (in this case, 2-body) localize the evolution.  Consistent with this result,  it can be seen in Fig. \ref{fig:IPR and G}a that the IPR approaches a constant value ($\sim 0.8$) when disorder is included.  In Fig. \ref{fig:IPR and G}b, the average density of spin-up particles \footnote{Due to the up-down symmetry of the evolution, the average density of spin-down particles is equivalent to the average density of spin-up particles and thus the spin-down density plots are not included.} at each site is plotted after $10,000$ Floquet periods.  Without disorder, the particle spreads throughout the entire system.  With disorder, it remains localized around its initial location. 

In the second case, we evolve under Hubbard-RLBL with the parameters $V=7.9$, $\tau = 2 \pi + 0.1$.  This point in parameter space is not in the vicinity of any Diophantine points, but, nonetheless, the first and fourth conditions (single particles frozen, doublons swapping) in \eqref{eq: freeze condition} are approximately satisfied (since the point is in an overlapping red and green region in Fig. \ref{fig:HubbardRLBL Phase Diagram}).  Similar to the first case, the evolution without disorder exhibits slowed, effective dynamics with the IPR decreasing over time (Fig. \ref{fig:IPR and G}a).  When disorder is added, the IPR again converges to a constant value ($\sim 0.8$).  Note, the Fock state permutation corresponding to single particles frozen and doublons swapping does not have finite order.  However, it does have finite orbits in the two particle subspace, and, furthermore, the evolution is $U(1)$ symmetric.  Thus, evolution is confined to the two particle subspace.  This implies that, consistent with the numerics, disorder is again expected to localize the system by the arguments of Section \ref{sec: subspaces}.  

In the third case, we consider evolution at $V=\sqrt{2} - 0.01$, $\tau = \sqrt{2} \pi + 0.01$.  Here, only the third condition in \eqref{eq: freeze condition} is approximately satisfied (doublons frozen).  The exact satisfaction of the third condition fragments the Hilbert space leaving any configuration of doublons frozen while other particle configurations may thermalize.  Since, in this case, the third condition is only approximately satisfied, doublons may separate into a spin-up and spin-down particle.  This means that, unlike case 2, the perturbation allows the system to evolve out of the Fock state permutation subspace and, as discussed at the end of Section \ref{sec: subspaces}, disorder is not guaranteed to localize the evolution. 
 This is reflected in Fig. \ref{fig:IPR and G}a where the IPR no longer converges once disorder is added.  However, the disorder does help to stabilize the frozen dynamics of the doublon over long time scales.  The difference from the first two cases is also apparent when considering the average particle density Fig. \ref{fig:IPR and G}b.  Here, the doublon is still with high probability localized near its initial location, but once it splits (thereby leaving the frozen subspace) the single particles may travel throughout the system.  This creates a non-zero background in the average particle density even far from the initial doublon location.

\section{Summary and discussion}
\label{sec: conclusion}

In this work, we have shown that interacting Floquet models with periodically activated pairs exhibit classical cellular automation dynamics corrected by a slowed, effective interacting evolution when a given set of conditions (e.g. \eqref{eq: freeze condition}) are approximately satisfied. 
 Furthermore, when disorder is added to the system, these regions with approximately satisfied conditions become robust prethermal phases.  If only a few of the Diophantine conditions are satisfied, the Hilbert space fragments into cellular automation subspaces and ergodic subspaces.  When the same conditions are instead approximately satisfied, the disorder stabilizes the dynamics in the cellular automation subspace for long, but not infinite, time scales.  On the other hand, these subspaces may still support localization if the subspace is closed under the evolution.

The existence of these stabilized cellular automation phases opens the door to a systematic investigation into their properties.  For example, in \cite{Nathan2021AFI} the disordered Hubbard-RLBL model was investigated at $\tau = \frac{\pi}{2}$ with $V$ approaching $0$ and $V$ approaching infinity.  It was found that, in this regime, the system belongs to a class of anomalous Floquet topological insulators, called correlation-induced anomalous Floquet insulators (CIAFI), labeled by a hierarchy of topological invariants.  The two different values of $V$ correspond to topological insulators with two differing topological invariants.  From the perspective of our work, this corresponds to the cases where single particles and doublons are approximately swapping (conditions two and four approximately satisfied) for $V$ near $0$ and corresponds to single particles swapping and doublons frozen (conditions two and three approximately satisfied) for $V$ large.  Thus, any parameter space region with those Diophantine conditions approximately satisfied will also correspond to a CIAFI with the corresponding topological invariants.  Similarly, when other Diophantine conditions are approximately satisfied, we expect the system to again correspond to a CIAFI with different topological invariants.  This is just one example of the interesting phenomena that may occur in systems with stabilized cellular automation dynamics and an exciting direction for future work is the investigation of possible exotic behavior found in systems stabilizing other cellular automata.  For example, it would be interesting to consider interacting Floquet drives without $U(1)$ symmetry such that the corresponding cellular automation may not preserve particle number.

One restriction used in this work to show localization was the finite order of the cellular automation.  It is an open question whether there are any systems where this constraint may be relaxed.  Another interesting possibility is the generalization of the models discussed in this work to aperiodically driven systems.  If this is possible, the prospective stabilized cellular automation corresponding to the evolution of such a drive would necessarily be aperiodic as well and therefore allow for more general stabilized cellular automata.  Recent work has suggested the existence of prethermal phases for aperiodically driven systems \cite{Zhao2021aperiodic,Zhao2022aperiodic}.    

Instead of periodic drives, it is also possible to restrict hopping to between pairs of sites via measurements.  Recently it was shown that, in this way, it is possible to mimic the RLBL procedure to produce protected edge transport alongside a local bulk via measurements alone \cite{wampler2022stirring}.  Due to the non-unitary nature of the measurements, the analysis in this paper does not directly apply in the measurement-induced setting.  A possible avenue for future investigations is determining if the stabilized cellular automation dynamics is also possible for measurement-induced systems and, if so, what similarities and differences does the dynamics have with the Floquet systems considered here.   

\emph{Acknowledgments.}
Our work was supported in part by the NSF grant DMR-1918207. IK thanks O. Motrunich for discussions.

\bibliography{ArrestedDevelopment.bib}

\begin{thebibliography}{10}

\bibitem{Rudner2020FTI}
Mark~S. Rudner and Netanel~H. Lindner.
\newblock Band structure engineering and non-equilibrium dynamics in floquet
  topological insulators.
\newblock {\em Nature Reviews Physics}, 2(5):229--244, May 2020.

\bibitem{Else2020DTCRev}
Dominic~V. Else, Christopher Monroe, Chetan Nayak, and Norman~Y. Yao.
\newblock Discrete time crystals.
\newblock {\em Annual Review of Condensed Matter Physics}, 11(1):467--499,
  2020.

\bibitem{Sacha2020DTCBook}
Krzysztof Sacha.
\newblock {\em Time Crystals}.
\newblock Springer Series on Atomic, Optical, and Plasma Physics. Springer
  Cham, 1 edition.

\bibitem{titum2016anomalous}
Paraj Titum, Erez Berg, Mark~S Rudner, Gil Refael, and Netanel~H Lindner.
\newblock Anomalous floquet-anderson insulator as a nonadiabatic quantized
  charge pump.
\newblock {\em Physical Review X}, 6(2):021013, 2016.

\bibitem{Nathan2019AFI}
Frederik Nathan, Dmitry Abanin, Erez Berg, Netanel~H. Lindner, and Mark~S.
  Rudner.
\newblock Anomalous floquet insulators.
\newblock {\em Phys. Rev. B}, 99:195133, May 2019.

\bibitem{Nathan2021AFI}
Frederik Nathan, Dmitry~A. Abanin, Netanel~H. Lindner, Erez Berg, and Mark~S.
  Rudner.
\newblock {Hierarchy of many-body invariants and quantized magnetization in
  anomalous Floquet insulators}.
\newblock {\em SciPost Phys.}, 10:128, 2021.

\bibitem{Wilczek2012TimeCrystal}
Frank Wilczek.
\newblock Quantum time crystals.
\newblock {\em Phys. Rev. Lett.}, 109:160401, Oct 2012.

\bibitem{watanabe2015notimecrystal}
Haruki Watanabe and Masaki Oshikawa.
\newblock Absence of quantum time crystals.
\newblock {\em Phys. Rev. Lett.}, 114:251603, Jun 2015.

\bibitem{Else2016DTC}
Dominic~V. Else, Bela Bauer, and Chetan Nayak.
\newblock Floquet time crystals.
\newblock {\em Phys. Rev. Lett.}, 117:090402, Aug 2016.

\bibitem{rudner2013anomalous}
Mark~S Rudner, Netanel~H Lindner, Erez Berg, and Michael Levin.
\newblock Anomalous edge states and the bulk-edge correspondence for
  periodically driven two-dimensional systems.
\newblock {\em Physical Review X}, 3(3):031005, 2013.

\bibitem{Zhang2017DTCExp}
J.~Zhang, P.~W. Hess, A.~Kyprianidis, P.~Becker, A.~Lee, J.~Smith, G.~Pagano,
  I.-D. Potirniche, A.~C. Potter, A.~Vishwanath, N.~Y. Yao, and C.~Monroe.
\newblock Observation of a discrete time crystal.
\newblock {\em Nature}, 543(7644):217--220, Mar 2017.

\bibitem{Frey2022DTCexp}
Philipp Frey and Stephan Rachel.
\newblock Realization of a discrete time crystal on 57 qubits of a quantum
  computer.
\newblock {\em Science Advances}, 8(9):eabm7652, 2022.

\bibitem{Peng2016AFAIExp}
Yu-Gui Peng, Cheng-Zhi Qin, De-Gang Zhao, Ya-Xi Shen, Xiang-Yuan Xu, Ming Bao,
  Han Jia, and Xue-Feng Zhu.
\newblock Experimental demonstration of anomalous floquet topological insulator
  for sound.
\newblock {\em Nature Communications}, 7(1):13368, Nov 2016.

\bibitem{Maczewsky2017AFAIExp}
Lukas~J. Maczewsky, Julia~M. Zeuner, Stefan Nolte, and Alexander Szameit.
\newblock Observation of photonic anomalous floquet topological insulators.
\newblock {\em Nature Communications}, 8(1):13756, Jan 2017.

\bibitem{Mukherjee2017AFAIExp}
Sebabrata Mukherjee, Alexander Spracklen, Manuel Valiente, Erika Andersson,
  Patrik {\"O}hberg, Nathan Goldman, and Robert~R. Thomson.
\newblock Experimental observation of anomalous topological edge modes in a
  slowly driven photonic lattice.
\newblock {\em Nature Communications}, 8(1):13918, Jan 2017.

\bibitem{Wintersperger2020AFAIExp}
Karen Wintersperger, Christoph Braun, F.~Nur {\"U}nal, Andr{\'e} Eckardt,
  Marco~Di Liberto, Nathan Goldman, Immanuel Bloch, and Monika Aidelsburger.
\newblock Realization of an anomalous floquet topological system with ultracold
  atoms.
\newblock {\em Nature Physics}, 16(10):1058--1063, Oct 2020.

\bibitem{Lazarides2014Therm}
Achilleas Lazarides, Arnab Das, and Roderich Moessner.
\newblock Equilibrium states of generic quantum systems subject to periodic
  driving.
\newblock {\em Phys. Rev. E}, 90:012110, Jul 2014.

\bibitem{Dalessio2014Therm}
Luca D'Alessio and Marcos Rigol.
\newblock Long-time behavior of isolated periodically driven interacting
  lattice systems.
\newblock {\em Phys. Rev. X}, 4:041048, Dec 2014.

\bibitem{Ponte2015Therm}
Pedro Ponte, Anushya Chandran, Z.~Papić, and Dmitry~A. Abanin.
\newblock Periodically driven ergodic and many-body localized quantum systems.
\newblock {\em Annals of Physics}, 353:196--204, 2015.

\bibitem{Dehghani2014Dissipative}
Hossein Dehghani, Takashi Oka, and Aditi Mitra.
\newblock Dissipative floquet topological systems.
\newblock {\em Phys. Rev. B}, 90:195429, Nov 2014.

\bibitem{Iadecola2015Bath}
Thomas Iadecola and Claudio Chamon.
\newblock Floquet systems coupled to particle reservoirs.
\newblock {\em Phys. Rev. B}, 91:184301, May 2015.

\bibitem{Iadecola2015Bath2}
Thomas Iadecola, Titus Neupert, and Claudio Chamon.
\newblock Occupation of topological floquet bands in open systems.
\newblock {\em Phys. Rev. B}, 91:235133, Jun 2015.

\bibitem{Seetharam2015Bath}
Karthik~I. Seetharam, Charles-Edouard Bardyn, Netanel~H. Lindner, Mark~S.
  Rudner, and Gil Refael.
\newblock Controlled population of floquet-bloch states via coupling to bose
  and fermi baths.
\newblock {\em Phys. Rev. X}, 5:041050, Dec 2015.

\bibitem{Canovi2016pretherm}
Elena Canovi, Marcus Kollar, and Martin Eckstein.
\newblock Stroboscopic prethermalization in weakly interacting periodically
  driven systems.
\newblock {\em Phys. Rev. E}, 93:012130, Jan 2016.

\bibitem{Mori2016pretherm}
Takashi Mori, Tomotaka Kuwahara, and Keiji Saito.
\newblock Rigorous bound on energy absorption and generic relaxation in
  periodically driven quantum systems.
\newblock {\em Phys. Rev. Lett.}, 116:120401, Mar 2016.

\bibitem{Kuwahara2016pretherm}
Tomotaka Kuwahara, Takashi Mori, and Keiji Saito.
\newblock Floquet–magnus theory and generic transient dynamics in
  periodically driven many-body quantum systems.
\newblock {\em Annals of Physics}, 367:96--124, 2016.

\bibitem{Abanin2017pretherm1}
Dmitry~A. Abanin, Wojciech De~Roeck, Wen~Wei Ho, and Fran\ifmmode
  \mbox{\c{c}}\else~\c{c}\fi{}ois Huveneers.
\newblock Effective hamiltonians, prethermalization, and slow energy absorption
  in periodically driven many-body systems.
\newblock {\em Phys. Rev. B}, 95:014112, Jan 2017.

\bibitem{Abanin2017pretherm2}
Dmitry Abanin, Wojciech De~Roeck, Wen~Wei Ho, and Fran{\c{c}}ois Huveneers.
\newblock A rigorous theory of many-body prethermalization for periodically
  driven and closed quantum systems.
\newblock {\em Communications in Mathematical Physics}, 354(3):809--827, Sep
  2017.

\bibitem{Abanin2019MBLRev}
Dmitry~A. Abanin, Ehud Altman, Immanuel Bloch, and Maksym Serbyn.
\newblock Colloquium: Many-body localization, thermalization, and entanglement.
\newblock {\em Rev. Mod. Phys.}, 91:021001, May 2019.

\bibitem{Fleishman1980MBL}
L.~Fleishman and P.~W. Anderson.
\newblock Interactions and the anderson transition.
\newblock {\em Phys. Rev. B}, 21:2366--2377, Mar 1980.

\bibitem{DAlessio2013MBL}
Luca D’Alessio and Anatoli Polkovnikov.
\newblock Many-body energy localization transition in periodically driven
  systems.
\newblock {\em Annals of Physics}, 333:19--33, 2013.

\bibitem{Ponte2015MBL}
Pedro Ponte, Anushya Chandran, Z.~Papić, and Dmitry~A. Abanin.
\newblock Periodically driven ergodic and many-body localized quantum systems.
\newblock {\em Annals of Physics}, 353:196--204, 2015.

\bibitem{Ponte2015MBLPRL}
Pedro Ponte, Z.~Papi\ifmmode~\acute{c}\else \'{c}\fi{}, Fran\ifmmode
  \mbox{\c{c}}\else~\c{c}\fi{}ois Huveneers, and Dmitry~A. Abanin.
\newblock Many-body localization in periodically driven systems.
\newblock {\em Phys. Rev. Lett.}, 114:140401, Apr 2015.

\bibitem{Lazarides2015MBL}
Achilleas Lazarides, Arnab Das, and Roderich Moessner.
\newblock Fate of many-body localization under periodic driving.
\newblock {\em Phys. Rev. Lett.}, 115:030402, Jul 2015.

\bibitem{Imbrie2016MBL}
John~Z. Imbrie.
\newblock On many-body localization for quantum spin chains.
\newblock {\em Journal of Statistical Physics}, 163(5):998--1048, Jun 2016.

\bibitem{Abanin2016MBLFloquet}
Dmitry~A. Abanin, Wojciech {De Roeck}, and François Huveneers.
\newblock Theory of many-body localization in periodically driven systems.
\newblock {\em Annals of Physics}, 372:1--11, 2016.

\bibitem{Khemani2016MBL}
Vedika Khemani, Achilleas Lazarides, Roderich Moessner, and S.~L. Sondhi.
\newblock Phase structure of driven quantum systems.
\newblock {\em Phys. Rev. Lett.}, 116:250401, Jun 2016.

\bibitem{Agarwala2017MBL}
Adhip Agarwala and Diptiman Sen.
\newblock Effects of interactions on periodically driven dynamically localized
  systems.
\newblock {\em Phys. Rev. B}, 95:014305, Jan 2017.

\bibitem{Anderson1958localize}
P.~W. Anderson.
\newblock Absence of diffusion in certain random lattices.
\newblock {\em Phys. Rev.}, 109:1492--1505, Mar 1958.

\bibitem{Sala2020HFrag}
Pablo Sala, Tibor Rakovszky, Ruben Verresen, Michael Knap, and Frank Pollmann.
\newblock Ergodicity breaking arising from hilbert space fragmentation in
  dipole-conserving hamiltonians.
\newblock {\em Phys. Rev. X}, 10:011047, Feb 2020.

\bibitem{Moudgalya2022Scars}
Sanjay Moudgalya, B~Andrei Bernevig, and Nicolas Regnault.
\newblock Quantum many-body scars and hilbert space fragmentation: a review of
  exact results.
\newblock {\em Reports on Progress in Physics}, 85(8):086501, jul 2022.

\bibitem{Motrunich22}
Sanjay Moudgalya and Olexei~I. Motrunich.
\newblock Hilbert space fragmentation and commutant algebras.
\newblock {\em Phys. Rev. X}, 12:011050, Mar 2022.

\bibitem{Turner2018Scars}
C.~J. Turner, A.~A. Michailidis, D.~A. Abanin, M.~Serbyn, and Z.~Papi{\'{c}}.
\newblock Weak ergodicity breaking from quantum many-body scars.
\newblock {\em Nature Physics}, 14(7):745--749, Jul 2018.

\bibitem{Ho2019Scars}
Wen~Wei Ho, Soonwon Choi, Hannes Pichler, and Mikhail~D. Lukin.
\newblock Periodic orbits, entanglement, and quantum many-body scars in
  constrained models: Matrix product state approach.
\newblock {\em Phys. Rev. Lett.}, 122:040603, Jan 2019.

\bibitem{Kumar2018evenodd}
Ajesh Kumar, Philipp~T. Dumitrescu, and Andrew~C. Potter.
\newblock String order parameters for one-dimensional floquet symmetry
  protected topological phases.
\newblock {\em Phys. Rev. B}, 97:224302, Jun 2018.

\bibitem{friedman2019integrable}
Aaron~J Friedman, Sarang Gopalakrishnan, and Romain Vasseur.
\newblock Integrable many-body quantum floquet-thouless pumps.
\newblock {\em Physical review letters}, 123(17):170603, 2019.

\bibitem{Ljubotina2019evenodd}
Marko Ljubotina, Lenart Zadnik, and Toma\ifmmode \check{z}\else~\v{z}\fi{}
  Prosen.
\newblock Ballistic spin transport in a periodically driven integrable quantum
  system.
\newblock {\em Phys. Rev. Lett.}, 122:150605, Apr 2019.

\bibitem{Piroli2020evenodd}
Lorenzo Piroli, Bruno Bertini, J.~Ignacio Cirac, and Toma\ifmmode
  \check{z}\else~\v{z}\fi{} Prosen.
\newblock Exact dynamics in dual-unitary quantum circuits.
\newblock {\em Phys. Rev. B}, 101:094304, Mar 2020.

\bibitem{Lu2022EvenOddPRL}
Mingwu Lu, G.~H. Reid, A.~R. Fritsch, A.~M. Pi\~neiro, and I.~B. Spielman.
\newblock Floquet engineering topological dirac bands.
\newblock {\em Phys. Rev. Lett.}, 129:040402, Jul 2022.

\bibitem{wampler2022arrested}
Matthew Wampler and Israel Klich.
\newblock Arrested development and fragmentation in strongly-interacting
  floquet systems.
\newblock {\em arXiv preprint arXiv:2209.09180}, 2022.

\bibitem{Cohen2007Dioph}
Henri Cohen.
\newblock {\em Number Theory}.
\newblock Graduate Texts in Mathematics. Springer, 1 edition.

\bibitem{Wolfram1983CA}
Stephen Wolfram.
\newblock Statistical mechanics of cellular automata.
\newblock {\em Rev. Mod. Phys.}, 55:601--644, Jul 1983.

\bibitem{Aizenman2009kbl}
Michael Aizenman and Simone Warzel.
\newblock Localization bounds for multiparticle systems.
\newblock {\em Communications in Mathematical Physics}, 290(3):903--934, Sep
  2009.

\bibitem{Note1}
A complex permutation matrix is a matrix where every row and column has a
  single non-zero element whose modulus is $1$.

\bibitem{Note2}
Note, for example, if $H_{dis}$ is included as a constant term throughout the
  evolution instead of only being applied during the $U_{dis}$ portion of the
  drive, then all our results still hold \cite {Nathan2021AFI}. In this case,
  the strength of the disorder during the first 4 steps of the floquet drive
  must be kept small. The strength during the $5^{th}$, disorder only, step may
  be made large by lengthening the time that $U_{dis}$ is applied.

\bibitem{Note3}
This is because \cite {Abanin2016MBLFloquet} uses a KAM type scheme to, order
  by order, find exactly the dressed $\ell $-bits of the Hamiltonian and thus
  prove it is MBL. This procedure may thus be stopped at some order $K$ to show
  K-body localization.

\bibitem{DeRoeck2017MBL}
Wojciech De~Roeck and Fran\ifmmode \mbox{\c{c}}\else~\c{c}\fi{}ois Huveneers.
\newblock Stability and instability towards delocalization in many-body
  localization systems.
\newblock {\em Phys. Rev. B}, 95:155129, Apr 2017.

\bibitem{DeRoeck2017Avalanche}
Wojciech De~Roeck and John~Z. Imbrie.
\newblock Many-body localization: stability and instability.
\newblock {\em Philosophical Transactions of the Royal Society A: Mathematical,
  Physical and Engineering Sciences}, 375(2108):20160422, 2017.

\bibitem{Potirniche2019Avalanche}
Ionut-Dragos Potirniche, Sumilan Banerjee, and Ehud Altman.
\newblock Exploration of the stability of many-body localization in $d>1$.
\newblock {\em Phys. Rev. B}, 99:205149, May 2019.

\bibitem{Morningstar2022Avalanche}
Alan Morningstar, Luis Colmenarez, Vedika Khemani, David~J. Luitz, and David~A.
  Huse.
\newblock Avalanches and many-body resonances in many-body localized systems.
\newblock {\em Phys. Rev. B}, 105:174205, May 2022.

\bibitem{nahum2018operator}
Adam Nahum, Sagar Vijay, and Jeongwan Haah.
\newblock Operator spreading in random unitary circuits.
\newblock {\em Physical Review X}, 8(2):021014, 2018.

\bibitem{Note4}
Due to the up-down symmetry of the evolution, the average density of spin-down
  particles is equivalent to the average density of spin-up particles and thus
  the spin-down density plots are not included.

\bibitem{Zhao2021aperiodic}
Hongzheng Zhao, Florian Mintert, Roderich Moessner, and Johannes Knolle.
\newblock Random multipolar driving: Tunably slow heating through spectral
  engineering.
\newblock {\em Phys. Rev. Lett.}, 126:040601, Jan 2021.

\bibitem{Zhao2022aperiodic}
Hongzheng Zhao, Mark~S. Rudner, Roderich Moessner, and Johannes Knolle.
\newblock Anomalous random multipolar driven insulators.
\newblock {\em Phys. Rev. B}, 105:245119, Jun 2022.

\bibitem{wampler2022stirring}
Matthew Wampler, Brian J.~J. Khor, Gil Refael, and Israel Klich.
\newblock Stirring by staring: Measurement-induced chirality.
\newblock {\em Phys. Rev. X}, 12:031031, Aug 2022.

\end{thebibliography}
\bibliographystyle{unsrt}
\onecolumngrid
\appendix

\section{Solutions of Few Site Subspaces}
\subsection{Hubbard Floquet Evolution of 2-site Pair in the 2-particle Sector}

We index the 4-particle configurations of the subspace as follows:

\begin{subequations}
\begin{gather}
    0 \rightarrow \hbox{ } \uparrow \downarrow \hbox{\space \space } \text{ \textunderscore \textunderscore}\\
    1 \rightarrow \hbox{ } \text{\textunderscore \textunderscore} \hbox{\space \space \space}\uparrow \downarrow\\
    2 \rightarrow \hbox{ } \uparrow\text{\textunderscore} \hbox{\space\space \space} \text{\textunderscore}\downarrow\\
    3 \rightarrow \hbox{ } \text{\textunderscore}\downarrow \hbox{ } \uparrow\text{\textunderscore}
\end{gather}
\end{subequations}
We therefore have that the representation of the Hubbard Hamiltonian in this subspace is given by

\begin{gather}
    {\cal H} = \left(\begin{tabular}{c c c c}
         $V$ & $0$ & $-1$ & $-1$ \\
         $0$ & $V$ & $-1$ & $-1$ \\
         $-1$ & $-1$ & $0$ & $0$ \\
         $-1$ & $-1$ & $0$ & $0$ 
    \end{tabular} \right)
\end{gather}
Hence, the evolution, ${\cal U} = e^{-i {\cal H} \tau}$, is given by

\begin{gather}
    {\cal U} = e^{-\frac{1}{2}i V \tau} \left(\begin{tabular}{c c c c}
         $e^{-\frac{1}{2}i V \tau}\left[\frac{1}{2} + A \right]$ & $e^{-\frac{1}{2}i V \tau}\left[-\frac{1}{2} + A \right]$ & $B$ & $B$ \\
         $e^{-\frac{1}{2}i V \tau}\left[-\frac{1}{2} + A \right]$ & $e^{-\frac{1}{2}i V \tau}\left[\frac{1}{2} + A \right]$ & $B$ & $B$ \\
         $B$ & $B$ & $e^{\frac{1}{2}i V \tau}\left[\frac{1}{2} + \Bar{A} \right]$ & $e^{\frac{1}{2}i V \tau}\left[-\frac{1}{2} + \Bar{A} \right]$ \\
         $B$ & $B$ & $e^{\frac{1}{2}i V \tau}\left[-\frac{1}{2} + \Bar{A} \right]$ & $e^{\frac{1}{2}i V \tau}\left[\frac{1}{2} + \Bar{A} \right]$
    \end{tabular} \right)
    \label{eq: U representation for 2 site hubbard}
\end{gather}
where

\begin{gather}
    A(V,\tau) = \frac{e^{\frac{1}{2} i V \tau}}{2} \left[\cos(\frac{1}{2}\tau \sqrt{16+V^2}) - i \frac{V}{ \sqrt{16+V^2}} \sin(\frac{1}{2}\tau \sqrt{16+V^2}) \right] \\
    B(V,\tau) = 2 i \frac{ \sin(\frac{1}{2} \tau \sqrt{16 + V^2})}{\sqrt{16+V^2}}.
\end{gather}
We are now interested in finding when \eqref{eq: U representation for 2 site hubbard} is a complex permutation matrix.  Note, for non-zero $V$, $|B|<1$.  Thus, our only hope for a permutation matrix is if $B=0$.  This occurs when $\frac{1}{2}\tau \sqrt{16+V^2} = \pi m$ for some $m \in \integers$.

Solving for $A(V,\tau)$ at $\frac{1}{2}\tau \sqrt{16+V^2} = \pi m$ yields

\begin{gather}
    A(V,\tau)|_{m ~ condition} = \frac{1}{2} e^{i [\pi m + \frac{1}{2} V \tau]} 
    \label{eq: A after cond 1}
\end{gather}
In \eqref{eq: U representation for 2 site hubbard}, $\cal U$ is a permutation matrix when, in addition to the requirement $B=0$, $|A|=\frac{1}{2}$ and  $\frac{V\tau}{\pi} \in \integers$.  These 2 conditions are uniquely met when, using \eqref{eq: A after cond 1}, $\pi m + \frac{1}{2}V \tau = \pi n$ for some $n \in \integers$.  Thus, solving $\pi m + \frac{1}{2}V \tau = \pi n$ and $\frac{1}{2}\tau \sqrt{16+V^2} = \pi m$ for $V$ and $\tau$, we get conditions three and four in \eqref{eq: freeze condition}.  Combining conditions three and four with \eqref{eq: U representation for 2 site hubbard}, $\cal U$ then becomes

\begin{gather}
    {\cal U}|_{\text{Conditions: 3 and 4}} = \left(\begin{tabular}{c c c c}
         $n-1$ & $n$ &$0$&$0$\\
         $n$ & $n-1$ &$0$&$0$\\ 
         $0$&$0$&$n-1$ & $n$\\
         $0$&$0$&$n$ & $n-1$\\ 
    \end{tabular}\right)\mod 2
\end{gather}
i.e. yielding the result that when $n$ is even (odd) evolution is the identity (perfect swapping).

\end{document}